\documentclass[12pt]{article}

\usepackage[body={17.5cm, 21cm},right=2cm]{geometry}

\usepackage{color}
\usepackage{graphicx}
\usepackage{hyperref}
\usepackage{epsf}
\usepackage{graphicx,epsfig}
\usepackage{amsmath}
\usepackage{amssymb}
\usepackage{epsfig}
\usepackage{cite}
\usepackage{color,colordvi}
\newcommand{\be}{\begin{eqnarray}}
\newcommand{\ee}{\end{eqnarray}}
\newcommand{\bi}{\begin{itemize}}
\newcommand{\ei}{\end{itemize}}

\newcounter{hran}

\def\MSbar{\relax\ifmmode\overline{\rm MS}\else{$\overline{\rm MS}${ }}\fi}
\def\del{\partial}

\def\simlt{\stackrel{<}{{}_\sim}}

\numberwithin{equation}{section}
\begin{document}
\vspace{5mm}
\vspace*{0.5cm}
\begin{center}

\def\thefootnote{\fnsymbol{footnote}}
\thispagestyle{empty}
{\Large \bf 
Comments on the Starobinsky Model \\
\vspace{0.25cm}
of Inflation and its Descendants
\vspace{0.25cm}	
}
\vskip.4in
{\large  Alex Kehagias  $^{a,b}$, Azadeh Moradinezhad Dizgah$^{a}$
and Antonio Riotto$^{a}$}
\\[0.5cm]

\vspace{.3cm}
{\normalsize { \it $^{a}$ Department of Theoretical Physics and Center for Astroparticle Physics (CAP)\\ 24 quai E. Ansermet, CH-1211 Geneva 4, Switzerland}}\\

\vspace{.3cm}
{\normalsize {\it  $^{b}$ Physics Division, National Technical University of Athens, \\15780 Zografou Campus, Athens, Greece}}\\

\vspace{.3cm}


\end{center}

\vspace{3cm}

\hrule \vspace{0.3cm}
{\small  \noindent \textbf{Abstract} \\[0.3cm]
\noindent 
We point out  that the ability  of some  models of inflation, such as Higgs inflation and the universal attractor models, in reproducing the available data  is due  to their relation to the Starobinsky model of inflation.   
For large field values,  where the inflationary  phase takes place, all these classes of models are indeed identical to the    Starobinsky model.  Nevertheless,  the inflaton  is just an auxiliary field in  the Jordan frame of the Starobinsky model and this leads to two important consequences: first, the inflationary predictions of the Starobinsky model and its descendants are  slightly different (albeit not measurably); secondly 
the theories   have different small-field behaviour,  leading to different ultra-violet cut-off scales.  In particular,  one interesting descendant of the Starobinsky model
is the non-minimally-coupled quadratic chaotic inflation. Although the standard quadratic  chaotic inflation is ruled out by the recent Planck data,  its   non-minimally coupled version is in agreement with observational data and valid up to Planckian scales.

\vspace{0.5cm}  \hrule
\vskip 1cm

\def\thefootnote{\arabic{footnote}}
\setcounter{footnote}{0}


\baselineskip= 15pt

\newpage 

\section{Introduction}\pagenumbering{arabic}
\noindent
 The recent Planck results  \cite{planck} have indicated that the cosmological perturbations in the Cosmic Microwave Background (CMB) radiation are nearly gaussian and of the adiabatic type. If one insists in assuming that these perturbations are to be ascribed to single-field models of inflation \cite{lr}, the data  put severes restriction on the inflationary parameters. In particular, the Planck results have strengthened the upper limits on the tensor-to-scalar ratio, $r\simlt 0.12$ at 95\% C.L., disfavouring  many inflationary models \cite{planck}. For instance,   
 the  chaotic models with potential $\phi^n$ with $n\geq 2$ are not in good shape; in particular, the simplest quadratic chaotic model $m^2\phi^2$ has been excluded at about $95\%$ CL. 
 
Among the inflationary models discussed by the Planck collaboration is the Starobinsky $(R+R^2)$ theory proposed in Ref. \cite{star}, whose  predictions for the perturbations were originally discussed in Ref.  \cite{MG}. Although this model looks quite ad hoc at the theoretical level, its perfect agreement with the Planck data is basically due to an additional $1/N$ suppression ($N$ being the number of e-folds till the end of inflation) of $r$ with respect to the prediction for the scalar spectral index $n_s$. As expected,  this has  renewed interest in this model. Particular recent efforts have  been in the direction of the  the supersymmetric version of it \cite{EON,KL,FKR,FKLP12, FKP,FFS,KS1}, along the lines
originated in Refs. \cite{cecotti,CFPS}.

Of course there are also other models which are  in agreement with the Planck data. For example, the so-called Higgs inflation \cite{s1,s2,higgs} and
the so-called  universal attractor models \cite{attractor,attractor2} give exactly the same inflationary predictions to leading order as the  Starobinsky theory. 
In this paper we stress that there is a simple reason why this apparent coincidence takes place: all these models are the Starobinsky model during inflation. While this might be known to some (see for instance Ref. \cite{bez} for the Higgs model of inflation), it seems to be mysterious to others  \cite{KL0}. In the Planck paper \cite{planck}, for instance, the Starobinsky and the Higgs inflation models are treated as  different. There reason why these models may be considered descendants of the Starobinsky model is that during inflation the  kinetic terms are sub-leading with respect to the potential terms and therefore they can be neglected  in 
first approximation. If so, the scalar field present in the Higgs model and in the universal attractor models is just an auxiliary field which can be integrated out, giving rise to the  Starobinsky model. During the  inflationary phase, where kinetic energies are negligible, apparent unrelated models are described effectively by the same dynamics. 

The next natural question is therefore if one can  distinguish these descendants from the Starobinsky model. An obvious way is to compare the inflationary parameters in these models  beyond the leading order. As we will show, the slow-roll parameters are the same up to $\sim 10^{-5}$ corrections, which are quite small to be measured in the upcoming measurements. Another difference relies on the different
way reheating after inflation proceeds in the different models \cite{bez}, but again differences  are of the order of $10^{-3}$ in the spectral index, hardly detectable by Planck (the often quoted  Planck result $n_s=0.960\pm 0.007$ is based on assumptions on the reionization, 
the primordial Helium abundance and
the effective number of neutrino). 

The fact that the Starobinsky model and its descendants differ by the kinetic term is also interesting from another point of view. While the kinetic terms play a sub-leading role during inflation, they play a fundamental role in determining the Ultra-Violet (UV) behavior of the theories and its cut-off $\Lambda$.  In particular, there is an ongoing discussion about the validity of the Higgs inflation as it seems that the cut-off of this theory is lower that the inflationary scale  \cite{burgess,barbon,hertz} (see Ref. \cite{mag} for a criticism to these results). 
On the other side, the cut-off of the Starobinsky theory is the Planck scale $M_{\rm p}$ \cite{hertz} so that inflation can be trusted in this framework.  The difference relies exactly in the role played by the kinetic energy. We will extend  the discussion of the cut-off for the universal attractor models. We will find that when the potential in Jordan frame is of the power-law type $\sim \phi^{2n}$, the cut-off is always above the inflationary scale  only for $n>7/2$. Therefore, for any value of $n<7/2$ (like for example the Higgs inflation case for which  $n=2$),  the cut-off  satisfies the relation $\Lambda< V^{1/4}$, where $V$ is the vacuum energy driving inflation, thus making the inflationary predictions questionable.  The case $n=1$ is particular as it corresponds to a non-minimally coupled simple quadratic chaotic inflation. We find in this case that the cut-off of this theory is at the Planck scale as in Starobinsky theory. Therefore, inflation can be trusted for the non-minimally coupled version of the simple quadratic chaotic inflation.  
 
The structure of this work is as follows. In section 2 we briefly describe the  Starobinsky model and show why the  Higgs inflation model, the  universal attractor models as well  as a higher-dimensional Starobinsky-like model, which is related to the $T$-model of Ref. \cite{KL0}, 
may be considered descendants of the Starobinsky model during inflation. In section 3 we discuss the differences between these models in their predictions for inflationary parameters, deferring the discussion of their cut-offs,  if viewed as effective field theories, until section 4. Finally,  we conclude in section 5.

 \section{The Starobinsky model and its descendants}
The Starobinsky model  \cite{star}  is described by the Lagrangian 

\be
 S_{\rm S}=\frac{1}{2}\int {\rm d}^4 x\sqrt{-g} \ \left(M_{\rm p}^2R+\frac{1}{6M^2} R^2\right).  \label{star}
 \ee
 This theory propagates a spin-2 state (graviton) and a scalar degree of freedom. The latter is manifest in the so-called linear representation
where one can  rewrites the Lagrangian (\ref{star}) as \cite{Whitt:1984pd}
\be
\label{R2}
 S_{\rm S}=\int {\rm d}^4 x\sqrt{-g} \ \left(\frac{M_{\rm p}^2}{2}R + \frac{1}{M}  R\psi-
 3 \psi^2\right).
 \ee
It is easy to see that  upon integrating out $\psi$, one gets back the original theory (\ref{star}).
After writing the expression (\ref{R2}) in the Einstein
frame by means of the conformal transformation
\be
g_{\mu\nu}\to e^{-\sqrt{2/3} \phi/M_{\rm p}}g_{\mu\nu}=\left(1+\frac{2\psi}{M M_{\rm p}^2}\right)^{-1} g_{\mu\nu}, 
\ee
we get the equivalent scalar field version of the Starobinsky model 
\be
 S_{\rm S}=\int {\rm d}^4 x\sqrt{-g}\left[\frac{M_{\rm p}^2}{2}R-\frac{1}{2}\partial_\mu\phi\partial^\mu \phi-\frac{3}{4} M_{\rm p}^4 M^2\left(1-e^{-\sqrt{\frac{2}{3}}\phi/M_{\rm p}}\right)^2\right]. \label{R3}
\ee
We see that during inflation (large values of $\phi$), 
the dynamics is dominate by the vacuum energy 
\be 
V_{\rm S}=\frac{3}{4}M_{\rm p}^4 M^2.  \label{VS}
\ee
Eq. (\ref{R3}) is the linear representation of the Starobinsky model where the extra scalar degree of freedom is manifest. The theory (\ref{R3}) leads to inflation with scalar tilt and  tensor-to-scalar ratio
\be
n_s-1\approx
-\frac{2}{N},\, \,\,r\approx \frac{12}{N^2} 
\ee
 Note that $r$ has an addition $1/N$ suppression  with respect to the scalar tilt and thus predicting a tiny amount of gravitational waves. It is therefore consistent with the Planck constraints. The normalization of the CMB anisotropies  fixes  $M\approx 10^{-5}$. 
 
\subsection{Higgs Inflation as a descendant of the Starobinsky model}
\noindent 
 Let us now consider Higgs inflation model which is described by an action of the form \cite{higgs}

 \be
 S_{\rm HI}=\int {\rm d}^4 x\sqrt{-g}\left[\frac{M_{\rm p}^2}{2}R+\xi H^\dagger H R-\partial_\mu H^\dagger \partial^\mu H 
 -\lambda(H^\dagger H-v^2)^2\right],
 \ee
 where $H$ is the SM Higgs doublet and $v$ its vacuum expectation value.
 In the unitary gauge $H=h/\sqrt{2}$ and for $h^2\gg v^2$ the theory is described by 

 \be
 S_{\rm HI}=\int {\rm d}^4 x\sqrt{-g}\left(\frac{M_{\rm p}^2}{2}R+\frac{1}{2}\xi h^2 R-\frac{1}{2}\partial_\mu h \partial^\mu h 
 -\frac{\lambda}{4}h^4\right). \label{HI}
 \ee
 In this case,  successful inflation exists for $\xi^2/\lambda \approx 10^{10}$.
 During inflation, the kinetic term is, by definition, smaller  than any potential term and thus 
 (\ref{HI}) is effectively described by the action
 
 \be
 S_{\rm HI}=\int {\rm d}^4 x\sqrt{-g}\left(\frac{M_{\rm p}^2}{2}R+\frac{1}{2}\xi h^2 R 
 -\frac{\lambda}{4}h^4\right).  \label{higgs-star}
 \ee
The Higgs field during inflation has been turned into an auxiliary field which can be integrated out. We find that
 \be
 \xi h R-\lambda h^3=0,
 \ee
 which leads to 
 \be
 h^2=\frac{\xi R}{\lambda}.
 \ee
  Plugging back this value 
 into the action, we find that the theory during inflation can equally well be described by
 \be
 S_{\rm HI}=\int {\rm d}^4 x\sqrt{-g}\left(\frac{M_{\rm p}^2}{2}R+\frac{\xi^2}{4\lambda} R^2\right).
 \ee
 Therefore, during inflation, Higgs inflation is the  Starobinsky model  (\ref{star}), one simple has to identify  
 \be
 M^2=\frac{\lambda}{3\xi^2}. \label{m}
 \ee
 Since we know that $M\approx 10^{-5}$, we get that $\xi^2\approx 10^{10} \lambda$, which 
  is, not surprisingly, the value needed in Higgs Inflation. In addition, the vacuum energy which drives inflation is then
  \be 
  V_{\rm HI}=\frac{3}{4}M^2M_{\rm p}^4=\frac{\lambda}{4\xi^2}M_{\rm p}^4\, . \label{VHI}
  \ee  

 \subsection{Universal attractor models as a descendant of the Starobinsky model}
\noindent 
 The equivalence of the Starobinsky and Higgs inflation models is not merely an accident. In fact, the  Starobinsky model is 
 also equivalent during inflation to the general form of non-minimal coupling proposed in Ref. \cite{attractor}
 \be
 S_{\rm att}=\int {\rm d}^4 x \sqrt{-g}\left[\frac{1}{2}\Omega(\phi)R-\frac{1}{2}\partial_\mu \phi\partial^\mu\phi
 -V_J(\phi)\right], \label{ac0}
 \ee
 with 
 \be
 \Omega(\phi)=M_{\rm p}^2+\xi f(\phi), ~~~V_J=f(\phi)^2.
 \ee
 Is should be noted that this kind of models has been discussed first  in Ref. \cite{barbon}  where it was pointed out that they are not  technically ``natural" as  s there is no obvious way, a symmetry for example, to preserve the relation between the non-minimal coupling and the scalar potential.   
 
 As in the Higgs inflation case, during inflation, the dynamics is completely dominated by the potential so that we may 
 ignore the scalar kinetic term. Therefore, the theory turns out to be written  as
 \be
 S_{\rm att}=\int {\rm d}^4 x \sqrt{-g}\left[\frac{M_{\rm p}^2}{2}R+\frac{1}{2}\xi f(\phi)R
 -f(\phi)^2\right]. \label{ac1}
 \ee
 We may integrate out the scalar through its  equation of motion which is 
 \be
 \frac{1}{2}\xi R f'-2 f'f=0, ~~~~f'=\partial f/\partial\phi.
 \ee
 The scalar field equation admits two solutions
 \be
 f'=0\label{f1}
 \ee
 and 
 \be
 f=\frac{1}{4}\xi R. \label{f2}
 \ee
 Eq. (\ref{f1}) is solved by a constant configuration $\phi=\phi_*$. Therefore, it corresponds to Einstein 
 gravity with Planck mass $M_{\rm p}^2+\xi f(\phi_*)$ and cosmological constant $\lambda^2f(\phi_*)^2$. However, the second
 solution (\ref{f2}) gives 
 \be
 S_{\rm att}=\int {\rm d}^4 x \sqrt{-g}\left[\frac{M_{\rm p}^2}{2}R+\frac{\xi^2}{16}R^2\right]
 \ee
 {\it i.e.} the Starobinsky model (\ref{star}) again with the identification 
 \be
 M^2=\frac{4}{3\xi^2}.
 \ee
 The vacuum energy that drives inflation turns out to be for in this case
\be 
  V_{\rm att}=\frac{3}{4}M^2M_{\rm p}^4=\frac{M_{\rm p}^4}{\xi^2}\, . \label{VATT}
  \ee

\subsection{Higher-Dimensional Starobinsky model descendants}
\noindent 
Let us  now discuss the higher-dimensional generalization of the Starobinsky model with the action of the form  
 \be
 S=\int {\rm d}^{d}x\sqrt{-g}\left( \frac{M_*^{d-2}}{2}{\cal R}+a {\cal R}^b\right), \label{Sd}
 \ee 
 where ${\cal R}$ is the $(4+d)$-dimensional Ricci scalar, $M_*$ is the 
 corresponding Planck mass and $a$ and $b$ are dimensionless parameters. This  higher-dimensional theory  can be linearized in the scalar curvature as usual by
 introducing an auxiliary field $\phi$
 
\be
 S=\int {\rm d}^{d}x\sqrt{-g}\left( \frac{M_*^{d-2}}{2}{\cal R}+w\phi^2 \,{\cal R}-
 \phi^{\frac{2b}{b-1}}\right), \label{Sd1}
 \ee
 where 
 \be 
 w =\frac{b}{b-1}\Big{(}(b-1)a\Big{)}^{\frac{1}{b}}.
 \ee
 By making the conformal transformation to the metric $ g_{\mu\nu}\to \Omega^2 g_{\mu\nu}$, where
 \be 
 \Omega^{d-2}=\left(1+\frac{2w\phi^2}{M_*^{d-2}}\right)^{-1},
 \ee 
we may write the action (\ref{Sd1}) as 

\begin{align}
 S=\int {\rm d}^{d}x\sqrt{-g}\left(\phantom{X^{X^{X^X}}}\hspace{-1cm}\right.& \frac{M_*^{d-2}}{2}{\cal R}
-\frac{1}{2}(d-1)(d-2)M_*^{d-2}(\del_\mu \log \Omega)^2-\nonumber \\&\left.
-V_0\Big{\{}(\Omega^{2-d}-1)\Omega^{\frac{(b-1)d}{b}}\Big{\}}^{\frac{b}{b-1}}  
 \right), \label{Sd2}
 \end{align}
where 
\be
V_0=\frac{M_*^{\frac{b(d-2)}{b-1}}}{(2w)^{\frac{b}{(b-1)}}}.
\ee
Clearly, in order to get a Starobinsky-like model, we need 
\be 
d-2=\frac{b-1}{b}d ~~~\mbox{or} ~~~b=\frac{d}{2}.
\ee
Then the action  (\ref{Sd2}) turns out to be

\begin{align}
 S=\int {\rm d}^{d}x\sqrt{-g}\left[\frac{M_*^{d-2}}{2}{\cal R}
-\frac{1}{2}(d-1)(d-2)M_*^{d-2}(\del_\mu \log \Omega)^2-
V_0\Big{(}1-\Omega^{d-2}\Big{)}^{\frac{d}{d-2}}  
 \right]. \label{Sd3}
 \end{align} 
After parametrizing $\Omega$ as 
\be 
\log \Omega  = -\frac{1}{\sqrt{(d-1)(d-2)}}\frac{\psi}{{M_*}^{(d-2)/2}},
\ee
we get that 
\begin{align}
 S=\int {\rm d}^{d}x\sqrt{-g}\left[\frac{M_*^{d-2}}{2}{\cal R}
-\frac{1}{2}\del_\mu \psi\del^\mu\psi-
V_0\Big{(}1-e^{-\sqrt{\frac{d-2}{d-1}}\frac{\psi}{{M_*}^{\frac{d-2}{2}}}}\Big{)}^{\frac{d}{d-2}}  
 \right]. \label{Sd4}
 \end{align} 
After a dimensional reduction in a $d\!-\!4$ torus $T^{d-4}$, we get the four-dimensional
action 
\begin{align}
 S=\int {\rm d}^{4}x\sqrt{-g}\left[\frac{M_{\rm p}^{2}}{2}R
-\frac{1}{2}\del_\mu \chi\del^\mu\chi-
V_0\Big{(}1-e^{-\sqrt{\frac{d-2}{d-1}}\frac{\chi}{{M_{\rm p}}}}\Big{)}^{\frac{d}{d-2}}  
 \right] \label{Sd4}
 \end{align}
 after identifying 
 \be
 \chi=V_{d-4}^{1/2}\psi, ~~~~~V_{d-4} M_*^{d-2}=M_{\rm p}^2,
 \ee
 where $V_{d-4}$ is the volume of $T^{d-4}$. We assume of course that the torus moduli or at least its volume modulus are stabilized. 
 The potential of this generalized Starobinsky model is of the general form
\be 
V=V_0\Big{(}1-e^{\alpha\frac{\phi}{M_{\rm p}}}\Big{)}^{\beta},
\ee 
which is a kind of $T$-model \cite{KL0}. For such a potential, it is straightforward to calculate the inflationary predictions. We find that 

\be 
&&n_s
\approx 1-\frac{2}{N}, \nonumber\\
&&r
\approx \frac{8}{\alpha^2 N^2},
\ee
where $1/N_0=\alpha \sqrt{2}$ and we have taken the limit  $N\gg N_0$.
In this limit, this is same  with  
 the $T$-model predictions \cite{KL0,KL00} as during inflation, $\beta$ can be absorbed, to leading order, by appropriate shift of $\phi$ .
%

We conclude this section with a comment on the conformally invariant  SO(1,1) two-field model of Ref. \cite{KL0} described by the Lagrangian
\be
{\cal L}=\sqrt{-g}\left[\frac{1}{2}\partial_\mu \chi \del^\mu \chi+\frac{\chi^2}{12}R
-\frac{1}{2}\del_\mu\phi\del ^\mu \phi+\frac{\phi^2}{12}R-\frac{\lambda}{4}(\phi^2-\chi^2)^2\right]. \label{ll1}
\ee
The field $\chi$ has a wrong kinetic term and it was called conformon in Ref. \cite{KL0}. 
Clearly the Lagrangian (\ref{ll1}) is invariant under SO(1,1), rotations of $(\phi,\chi)$. Therefore, one may  fix this symmetry either by going to the Einstein frame  
$\chi^2-\phi^2=6M_{\rm p}^2$ or to the Jordan frame $\chi=\sqrt{6} M_{\rm p}$. Both gauge fixings lead to 
\be 
{\cal L}=\sqrt{-g}\left(\frac{M_{\rm p}^2}{2}R
-\frac{1}{2}\del_\mu\phi\del ^\mu \phi-9 \lambda\, M_{\rm p}^4\right). \label{c1}
\ee
Here, we will ignore as we did above the kinetic terms assuming that they are small compared to the potential term. In this case, $\phi$ and $\chi$ are auxiliaries which can be integrated out to give 
\be 
{\cal L}=\sqrt{-g}\frac{1}{144\lambda}\, R^2. \label{r2}
\ee
This is nothing else than Starobinsky model in the $M_{\rm p}\to \infty$ limit. Therefore, again the  conformally invariant SO(1,1) symmetric two-field model is a particular limit of the Starobinsky theory, at least in the region where scalar kinetic terms can be ignored. Note that (\ref{r2}) propagates a graviton and a scalar as can be seen in the linear representation 
\be 
{\cal L}=\sqrt{-g}\big{(}\varphi R-36 \lambda\varphi^2\big{)}. \label{r22}
\ee
By integrating out $\varphi$ we get the $R^2$ theory in (\ref{r2}).
By going to the Einstein frame by means of the conformal transformation 

\be
g_{\mu\nu}\to \frac{M_{\rm p}^2}{2\varphi}g_{\mu\nu}
\ee
we get 
\be 
{\cal L}=\sqrt{-g}\left(\frac{M_{\rm p}^2}{2}R
-\frac{3}{2\varphi^2}\del_\mu\varphi\del ^\mu \varphi-9 \lambda\, M_{\rm p}^4\right)
\ee
which is (\ref{c1}) after the transformation $\varphi=e^{\phi/\sqrt{3}}$.

 \section{Distinguishing Starobinsky model from its descendants}
From the discussion in the previous section, one can  conclude that the Starobinsky model and its descendants differ only in their kinetic terms. Therefore a reasonable question to ask is to which level this difference may be appreciated in the observables. Since the first slow-roll parameter
 $\epsilon=-\dot H/H^2$ (where $H$ is the Hubble rate during inflation) parametrizes the kinetic energy \cite{lr}, it is expected
 that differences between the Starobinsky model and its descendants appear at the level of differences in the slow-roll parameter $\epsilon$. For the Starobinsky model the slow-roll parameters are given by
 \be 
&&\epsilon_{\rm S}
\approx- \frac{3}{4N^2}, \\
&& \eta_{\rm S}
\approx -\frac{1}{N}.
\ee

%
Now let us consider the Higgs inflation model and re-write it in the Einstein frame. Redefining the metric as
 \be
 g_{\mu\nu}\to \left(1+\xi\frac{h^2}{M_{\rm p}^2}\right)^{-1}g_{\mu\nu},
 \ee
the action turns out to be
 \be
 S_{\rm HI}=
 \int {\rm d}^4 x\sqrt{-g}\left\{\frac{M_{\rm p}^2}{2}R-\frac{1}{2}\left(\frac{1}{1+\xi\frac{h^2}{M_{\rm p}^2}}+
 6\xi^2\frac{h^2}{M_{\rm p}^2}\frac{1}{\left(1+\xi\frac{h^2}{M_{\rm p}^2}\right)^2}\right)
 {\partial_\mu h \partial^\mu h }
 -\frac{\lambda}{4}\frac{h^4}{(1+\frac{\xi h^2}{M_{\rm p}^2})^2} \right\}. \label{hi1}
 \ee
 Let us now compare this theory with Starobinsky theory in the representation 
 (\ref{higgs-star}) which in the Einstein frame is written similarly as 
 \be
 S_{\rm S}=\int {\rm d}^4 x\sqrt{-g}\left(\frac{M_{\rm p}^2}{2}R-\frac{6}{2}\xi^2\frac{h^2}{M_{\rm p}^2}\frac{1}{\left(1+\xi\frac{h^2}{M_{\rm p}^2}\right)^2}
 {\partial_\mu h \partial^\mu h }
 -\frac{\lambda}{4}\frac{h^4}{(1+\frac{\xi h^2}{M_{\rm p}^2})^2} \right). \label{hi2}
 \ee
  The difference between the two theories is evident. They differ by a factor 
 \be
 \Delta {\cal L}=-\frac{1}{2}\frac{1}{1+\xi\frac{h^2}{M_{\rm p}^2}}
 {\partial_\mu h \partial^\mu h },  \label{dif-higgs}
 \ee
 which is precisely the Higgs kinetic term we neglected to arrive at the  Starobinsky theory
 in the Einstein frame.  
Here we should stress that the fundamental difference between the Higgs inflation and the Starobinsky model resides in the scalar kinetic term 
in the Jordan frame.
 For 
the Starobinsky model, there is no kinetic term for the auxiliary field $\phi$ in the linear representation of the model. This has the effect of making the parameter $\xi$ irrelevant as it can be completely absorbed in the scalar field and it is redundant. In the case of Higgs inflation there is a kinetic term for the Higgs field to start with, as it is a real field in Jordan frame and not an auxiliary. In this case therefore, $\xi$ cannot anymore be absorbed, it is not redundant and, as we shall see, 
it lowers the cut-off by a factor $\xi^{-1}$ as compared to Starobinsky model.   
 
 The slow-roll parameters for Higgs inflation and the Starobinsky theory are given by
 \be
&& \epsilon_{\rm HI,S}=\frac{M_{\rm p}^2}{2}\left(\frac{1}{V}\frac{\partial V}{\partial \chi}\right)^2=
 \frac{M_{\rm p}^2}{2}\left(\frac{1}{V}\frac{\partial V}{\partial h}\right)^2\left(\frac{\partial h}{\partial \chi}\right)^2,
 \ee
 where $\chi$ is the canonically normalized scalar, different for Higgs and Starobinsky models,  and $V$ is the common
 potential 
 \be
 V=\frac{\lambda}{4}\frac{h^4}{\left(1+\frac{\xi h^2}{M_{\rm p}^2}\right)^2}. 
 \ee
 Then, since  
 \be
\frac{\partial h}{\partial \chi}= \left(\frac{1}{1+\xi\frac{h^2}{M_{\rm p}^2}}+
 6\xi^2\frac{h^2}{M_{\rm p}^2}\frac{1}{\left(1+\xi\frac{h^2}{M_{\rm p}^2}\right)^2}\right)^{-1/2}
\ee
for Higgs inflation and 
\be
\frac{\partial h}{\partial \chi}= \left(6\xi^2\frac{h^2}{M_{\rm p}^2}\frac{1}{\left(1+\xi\frac{h^2}{M_{\rm p}^2}\right)^2}\right)^{-1/2}
\ee
for the Starobinsky model, we find that 
\be
&& \epsilon_{\rm HI}=
 \frac{M_{\rm p}^2}{2}\left(\frac{1}{V}\frac{\partial V}{\partial h}\right)^2 \left(\frac{1}{1+\xi\frac{h^2}{M_{\rm p}^2}}+
 6\xi^2\frac{h^2}{M_{\rm p}^2}\frac{1}{\left(1+\xi\frac{h^2}{M_{\rm p}^2}\right)^2}\right)^{-1},\\
 &&\epsilon_{\rm S}=
 \frac{M_{\rm p}^2}{2}\left(\frac{1}{V}\frac{\partial V}{\partial h}\right)^2 
 \left(6\xi^2\frac{h^2}{M_{\rm p}^2}\frac{1}{\left(1+\xi\frac{h^2}{M_{\rm p}^2}\right)^2}\right)^{-1}.\\
 \ee
   Since, the number of e-folds till the end of inflation  is related to $h$ as $
 N\approx (6\xi h^2/8 M_{\rm p}^2)$, 
 we get that 
 \be
 \frac{\epsilon_{\rm HI}}{\epsilon_{\rm S}}=\frac{8N \xi}{1+\frac{4}{3} N+8 N \xi}=1-\frac{1}{6\xi}\simeq 1-\frac{10^{-5}}{6\lambda}.
 \ee
Even though the slow-roll parameter enters with a factor of $6\epsilon$ in the spectral index $n_s$ , the difference is too small to be detectable. Another difference between the Starobinsky and the Higgs inflation model is their corresponding reheating temperatures \cite{bez}: $T_{\rm RH}\simeq 3\cdot 10^9$ GeV and $T_{\rm RH}\simeq 6\cdot 10^{13}$ GeV, respectively. This leads to a difference in the predicted value of spectral index at the level of $10^{-3}$ \cite{bez}. As we mentioned in the introduction, this difference is larger than the typical Planck error only if strong assumptions are made about the reionization history, 
the primordial Helium abundance and the effective number of neutrino. 

Let us now turn to the universal attractor models. The general class of models (\ref{ac0})  can be written in the Einstein frame by the conformal transformation
\be 
g_{\mu\nu}\to \left(1+\frac{\xi f(\phi)}{M_{\rm p}^2}\right)^{-2} g_{\mu\nu}
\ee
and it is explicitly written as
\be 
S_{\rm att}=\int {\rm d}^4 x \sqrt{-g}\left[\frac{M_{\rm p}^2}{2}R-\frac{3}{4}\frac{\xi^2f'^2}{M_{\rm p}^2}\frac{\del_\mu \phi\del^\mu \phi}{(1+\frac{\xi f}{M_{\rm p}^2})^2}  -\frac{1}{2}\frac{\del_\mu \phi\del^\mu \phi }{1+\frac{\xi f}{M_{\rm p}^2}}
 -\frac{f^2}{(1+\frac{\xi f}{M_{\rm p}^2})^2}\right]. \label{ac2}
 \ee
 Similarly, the  Starobinksy model in the representation (\ref{ac1}) can be  be written as 
 \be
 S_{\rm S}=\int {\rm d}^4 x \sqrt{-g}\left[\frac{M_{\rm p}^2}{2}R-\frac{3}{4}\frac{\xi^2}{M_{\rm p}^2}{f'}^2\frac{\del_\mu \phi\del^\mu \phi}{(1+\frac{\xi f}{M_{\rm p}^2})^2}  
 -\frac{f^2}{(1+\frac{\xi f}{M_{\rm p}^2})^2}\right]. \label{ac22}
 \ee
  Clearly, the two models differ in their kinetic terms 
  \be
  \Delta {\cal{L}}= -\frac{1}{2}\sqrt{-g} 
  \frac{\del_\mu \phi\del^\mu \phi }{1+\frac{\xi f}{M_{\rm p}^2}}.
  \label{diff-gen}
   \ee    
The slow-roll parameters for the above general classes of inflation models and the Starobinsky theory are given by
 \be
&& \epsilon_{\rm att, S}=\frac{M_{\rm p}^2}{2}\left(\frac{1}{V}\frac{\partial V}{\partial \chi}\right)^2=
 \frac{M_{\rm p}^2}{2}\left(\frac{1}{V}\frac{\partial V}{\partial \phi}\right)^2\left(\frac{\partial \phi}{\partial \chi}\right)^2,
 \ee
 where $\chi$ is the canonically normalized scalar, different for the two models and $V$ is the common
 potential 
 \be
 V=\frac{f^2}{\left(1+\frac{\xi f}{M_{\rm p}^2}\right)^2} .
 \ee
Let us discuss  the particular, but sufficiently generic case of  $f=\phi^n/M_{\rm p}^{n-2}$, for which
\be 
V = \frac{\phi^{2n}}{M_{\rm p}^{2n-4}(1+\xi\frac{\phi^n}{M_{\rm p}^n})}.
 \ee
 Then, since  
 \be
\frac{\partial \phi}{\partial \chi}= \left(\frac{1}{1+\xi\frac{\phi^n}{M_{\rm p}^n}}+
 \frac{3\xi^2n^2}{2}\frac{\phi^{2n-2}}{M_{\rm p}^{2n-2}}\frac{1}{\left(1+\xi\frac{\phi^n}{M_{\rm p}^n}\right)^2}\right)^{-1/2}
\ee
for general models of non-minimally coupled inflation and 
\be
\frac{\partial \phi}{\partial \chi}= \left(\frac{3\xi^2n^2}{2}\frac{\phi^{2n-2}}{M_{\rm p}^{2n-4}}\frac{1}{\left(1+\xi\frac{\phi^n}{M_{\rm p}^n}\right)^2}\right)^{-1/2}
\ee
for Starobinsky model, we find that 
\be
&& \epsilon_{\rm att}=
 \frac{M_{\rm p}^2}{2}\left(\frac{1}{V}\frac{\partial V}{\partial \phi}\right)^2  \left(\frac{1}{1+\xi\frac{\phi^n}{M_{\rm p}^n}}+
 \frac{3\xi^2n^2}{2}\frac{\phi^{2n-2}}{M_{\rm p}^{2n-2}}\frac{1}{\left(1+\xi\frac{\phi^n}{M_{\rm p}^n}\right)^2}\right)^{-1}\\
 &&\epsilon_{\rm S}=
 \frac{M_{\rm p}^2}{2}\left(\frac{1}{V}\frac{\partial V}{\partial \phi}\right)^2 
 \left(\frac{3\xi^2n^2}{2}\frac{\phi^{2n-2}}{M_{\rm p}^{2n-2}}\frac{1}{\left(1+\xi\frac{\phi^n}{M_{\rm p}^n}\right)^2}\right)^{-1}.\\
 \ee
 Since, the number of e-folding is related to $\phi$ as
 \be
 N \approx \frac{3 \xi \phi^n}{4M_{\rm p}^n} 
 \ee
 we infer that 
 \be
\frac{\epsilon_{\rm att}}{\epsilon_{\rm S}} \approx 1- \frac{N^{\frac{2}{n}-1}}{2n^2\xi^{\frac{2}{n}}}\left(\frac{4}{3}\right)^{2/n}.
\ee
This  always deviates from unity by a quantity smaller that $10^{-3}$ and therefore the difference is not observable.
 
 \section{Effective cut-off scales}
 One (somewhat  controversial) issue is the natural cut-off of the theories we have discussed so far. As there  exists another mass $M$ (or $1/\xi^{1/2}$), which enters besides the dimensionful Planck mass $M_{\rm p}$,   it is natural to expect that the cut-off of the theory may not be $M_{\rm p}$,  but 
 a ratio of it by appropriate power of $M$ (or $\xi$). If this power is high enough, it may happen that the cut-off is quit low, lower than the inflationary scale. In such a case, the discussion of inflation cannot be trusted or 
 it is questionable, to say the least. Bellow we will find the cut-offs of the models
 discussed so far by considering the scalar field in the Einstein frame as a 
 one-dimensional $\sigma$-model. Then, as mentioned, the expansion of its kinetic term  for small values of the field reveals  the cut-off of the theory and, above all, the differences among the models.

The  Starobinsky model (\ref{hi2}) can be expanded as 

\be 
S=\int {\rm d}^4 x\sqrt{-g}\left[\frac{M_{\rm p}^2}{2}R-\frac{1}{2}\left(\frac{\xi h^2}{M_{\rm p}^2}
+6\frac{\xi^2 h^2}{M_{\rm p}^2}+\cdots\right)\del_\mu h\del^\mu h-\frac{\lambda}{4}
h^4\left(1-2\frac{\xi h^2}{M_{\rm p}^2}+\cdots\right)\right]. \label{hi5}
\ee
We should canonically normalize  the leading kinetic term. Thus,  after defining $h^2=\frac{M_{\rm p}\psi}{\sqrt{3}\xi}$, 
 we get that the action  turns out to be
 \be
 S=\int {\rm d}^4 x\sqrt{-g}\left[\frac{M_{\rm p}^2}{2}R-\frac{1}{2}\left(1-\frac{ \psi}{\sqrt{3}M_{\rm p}}
+\cdots\right)\del_\mu \psi\del^\mu \psi-\frac{\lambda}{12}\frac{M_{\rm p}^2}{\xi^2}
\psi^2\left(1-2\frac{\psi}{\sqrt{3}M_{\rm p}}+\cdots\right)\right]. \label{hi4}
\ee
From the above form of the action we see that the cut-off $\Lambda_{\rm S}$ of  the Starobinsky theory is, as already found in \cite{hertz}
\be
\Lambda_{\rm S}=M_{\rm p}\, .
\ee 
A simple inspection of Eq. (\ref{VS}) 
shows that 
\be 
V_{\rm S}\ll \Lambda_{\rm S}^4,
\ee
indicating the internal consistency of the model \cite{hertz}.
The Higgs inflation action (\ref{hi1}) on the other hand can be expanded as 
\be 
S_{\rm HI}=\int {\rm d}^4 x\sqrt{-g}\left[\frac{M_{\rm p}^2}{2}R-\frac{1}{2}\left(1+\frac{\xi h^2}{M_{\rm p}^2}
+6\frac{\xi^2 h^2}{M_{\rm p}^2}+\cdots\right)\del_\mu h\del^\mu h-\frac{\lambda}{4}
h^4\left(1-2\frac{\xi h^2}{M_{\rm p}^2}+\cdots\right)\right]. \label{hi5}
\ee
Here the leading kinetic term is canonically normalized and therefore, since 
$\xi\gg 1$, we find that 
 the cut-off is \cite{barbon,burgess,hertz}
 \be 
 \Lambda_{\rm HI}=\frac{M_{\rm p}}{\xi}.
 \ee
This should be compared with the vacuum energy that drives inflation Eq. (\ref{VHI}), from where we get that  
\be 
V_{\rm HI}\gg \Lambda_{\rm HI}^4,
\ee
making the consistency of the model questionable.   This simple argument has been criticized in Ref. \cite{mag} where it was observed that
the cut-off should be field-dependent as the kinetic term is non-canonical. This argument would give a cut-off that during inflation, when $h\gg M_{\rm p}/\xi^{1/2}$, is even larger then the Planckian scale. However, we disagree with this approach. The presence of a cut-off $\Lambda_{\rm HI}\sim M_{\rm p}/\xi$ at lower values of the field cannot be avoided and it signals the breakdown of the model in that field range. 
The small field region is ``tested" by the dynamics during the reheating stage and one may not simply disregard this point by invoking that the inflationary field range is the one of interest. It should also be mentioned here a related problem,  the naturalness of the model.
The only way to solve this inconsistency is to add new degrees of freedom at energies $\sim M_{\rm p}/\xi^{1/2}$ in a way that does not  
spoil the flatness of the inflaton potential, as for example in the model discussed in Ref. \cite{GL}. In such a case, however,  predictability of  Higgs inflation is lost as there is now a strong dependence on the new physics assumed to appear at   $\sim M_{\rm p}/\xi^{1/2}$.

Similar considerations can be made for the attractor models. To find the cut-off $\Lambda_{\rm att}$, we  expand the action  (\ref{ac2}) as
 \begin{eqnarray} 
S_{\rm att}&\approx& \int {\rm d}^4 x \sqrt{-g}\left\{\frac{M_{\rm p}^2}{2}R-\frac{1}{2}
\left[1-\frac{\xi f}{M_{\rm p}}+\frac{\xi^2 f^2}{M_{\rm p}^2}+\cdots  3 \frac{\xi^2{f'}^2}{M_{\rm p}^2}\left(1-
2\frac{\xi f}{M_{\rm p}^2}+\frac{\xi^2{f'}^2}{M_{\rm p}^4}+\cdots\right)\right]\del_\mu \phi\del^\mu \phi \right.
 \nonumber \\
 &-&\left.f^2\left(1-2\xi f+\cdots\right)\right\}.\label{ac3}
 \end{eqnarray}
 For a polynomial form  $f(\phi)=\phi^n/M_{\rm p}^{n-2}$, with $n\neq 1$, we get that cut-off $\Lambda_{\rm att}$ is determined by the $\xi^2{f'}^2$ term in Eq. (\ref{ac3}) and reads 
 \be 
 \Lambda_{\rm att}=\frac{M_{\rm p}}{\xi^{\frac{1}{n-1}}}. \label{ct}
 \ee
Moreover, the vacuum energy during inflation is given in Eq. (\ref{VATT}), 
which in terms of the cut-off  (\ref{ct}) is written as
\be 
V_{\rm att}=
\xi^{\frac{6-2n}{n-1}}\Lambda_{\rm att}^4.
\ee
Clearly,  only for $n>7/3$ the vacuum energy satisfies $V_{\rm att}\ll \Lambda_{\rm att}^4 $ and the model makes sense. 
 
 The case $n=1$ is special and we will consider it separately. The reason is that $\xi^2{f'}^2$ dominates and a constant rescaling of the scalar, similarly to the one in the Starobinsky model,  is needed to canonically normalize the leading kinetic term.  
   It is known that the simplest chaotic inflation has severe problems with the recent Planck data. 
Its inflationary dynamics is described by the action
\be
S_m=\int {\rm d}^4 x \sqrt{-g}\left(\frac{M_{\rm p}^2}{2} R-\frac{1}{2}\del_\mu \phi \del ^\mu \phi-\frac{1}{2}m^2 \phi^2\right),
\label{mc}
\ee
which predicts \cite{lr} 
\be
n_s-1=-\frac{2}{N},~~~r=\frac{8}{N}, 
\ee
for the primordial tilt $n_s$ and the tensor-to-scalar ratio $r$, values which 
lie outside the joint 95\% CL for the Planck data. 
  Let us now consider instead of the action (\ref{mc}), a non-minimally coupled chaotic model 
 \be
 S=\int {\rm d}^4 x \sqrt{-g}\left(\frac{M_{\rm p}^2}{2} R+\xi M_{\rm p} \phi R
 -\frac{1}{2}\del_\mu \phi \del ^\mu \phi-\frac{1}{2}m^2 \phi^2\right),
\label{nmc}
\ee
 where $\xi$ is a dimensionless parameter. Clearly, as discussed above, during inflation
 the inflaton kinetic term is small compared to the potential and thus the model is described effectively by
 \be
 S=\int {\rm d}^4 x \sqrt{-g}\left(\frac{M_{\rm p}^2}{2} R+\xi M_{\rm p}\phi R
 -\frac{1}{2}m^2 \phi^2\right).
\label{nmc}
\ee
The field  $\phi$ can be integrated out leading again to Starobinsky model
\be
S=\int {\rm d}^4 x \sqrt{-g}\left(\frac{M_{\rm p}^2}{2} R+\xi^2\frac{M_{\rm p}^2}{2 m^2}R^2\right).
 \ee
 Therefore, the non-minimally coupled chaotic inflationary model (\ref{nmc}) is equivalent during inflation to the Starobinsky gravity (\ref{star}) with $M_{\rm p}^2/12M^2=M_{\rm p}^2\xi^2/2 m^2$. As a result,  since $M\approx 10^{-5}$,  we get that 
 \be
 \xi \approx 10^5 m,
 \ee
 whereas the primordial tilt and the tensor-to-scalar ratio are now $
(n_s-1)\simeq -2/N$ and $r=12/N^2$.
  Let us now write the action (\ref{nmc}) in the  Einstein frame. For this, we need to make the following conformal tranformation
 \be
 g_{\mu\nu}\to \left(1+\frac{2
\xi\phi}{M_{\rm p}}\right)^{-1}g_{\mu\nu}
\ee 
and the action becomes
\be 
S=\int {\rm d}^4 x \sqrt{-g}\left[\frac{M_{\rm p}^2}{2} R-3\xi^2\frac{\del_\mu\phi\del^\mu\phi}{\big{(}1+\frac{2\xi\phi}{M_{\rm p}}\big{)}^2}-\frac{1}{2}\frac{\del_\mu\phi\del^\mu\phi}{1+\frac{2\xi\phi}{M_{\rm p}}}-\frac{1}{2}m^2\phi^2\left(1+\frac{2\xi\phi}{M_{\rm p}}\right)^{-2}
\right].\label{nmc1}
\ee
%
For large values of the scalar field $\phi$ ($\phi \gg M_{\rm p}/2\xi$), we have
\be 
S_{nm}\approx \int {\rm d}^4 x \sqrt{-g}\left\{\frac{M_{\rm p}^2}{2} R-\frac{1}{2}\left(\frac{3M_{\rm p}^2}{2\phi^2}+\frac{M_{\rm p}}{2\xi \phi}\right)\del_\mu\phi\del^\mu\phi-V_0\left(1-\frac{M_{\rm p}}{\xi \phi}+\cdots\right)
\right\}\label{nmc3}
\ee
where 
\be
V_0=\frac{m^2M_{\rm p}^2}{8 \xi^2} \label{v0}
\ee
is the vacuum energy driving inflation.
Then, one may easily verify that this theory is Starobinksy theory for 
\be 
\frac{3M_{\rm p}^2}{2\phi^2}\gg\frac{M_{\rm p}}{2\xi \phi}
\ee
In other words, for $\phi$ in the range 
\be 
\frac{M_{\rm p}}{2\xi}\ll\phi\ll\frac{3}{2}\xi M_{\rm p} \label{cond}
\ee
the non-minimal chaotic inflation effectively coincides with Starobinsky model. 
Note that (\ref{cond}) implies that $\xi\gg1$. 

The action (\ref{nmc1}) can be expanded also for small values of $\phi$. 
However, in this case as  there is no canonically normalized leading  order kinetic term for the scalar. Thus, after defining $\chi=\sqrt{6}\xi \phi$, we have 
 \begin{eqnarray} 
S&\approx& \int {\rm d}^4 x \sqrt{-g}\left\{\frac{M_{\rm p}^2}{2} R-\frac{1}{2}\left(1-\frac{4\chi}{\sqrt{6}M_{\rm p}}-\frac{\chi}{3\xi M_{\rm p}}\right) \del_\mu\chi\del^\mu\chi\right.\nonumber \\
&-&\left.-\frac{1}{12}\frac{m^2M_{\rm p}^2}{\xi^2}\chi^2
\left(1-\frac{4\chi}{\sqrt{6}M_{\rm p}}\right)+\cdots
\right\}.\label{nmc4}
\end{eqnarray} 
From the form of the action, it follows  that the cut-off $\Lambda$  of the non-minimal chaotic inflation is   indeed the Planckian mass, $\Lambda=M_{\rm p}$, with $V_0\ll \Lambda^4$.  This is exactly what happens in the Starobinsky theory, where the absence of the canonically normalized leading kinetic term pushes the cut-off to the Planck scale.

\section{Conclusions}
\noindent
In this paper we have discussed the relation of certain inflationary models to the Starobinsky theory. In particular, we have pointed out that the agreement of these models with the recent Planck measurements in due to the fact that during inflation, they are effectively described by the Starobinsky theory. In this respect, the Starobinsky theory is a prototype of theories where the scalar potential has a plateau for large values of the scalar field. The examples we discussed here in details are the Higgs inflation model and the universal attractor models, the dynamics of which coincides to leading order in the slow-roll parameter with that of the Starobinsky theory. However, they differ from  the latter since the scalar  in the Starobinsky theory  is auxiliary in the Jordan frame and turns out to be propagating only in the Einstein frame. 

Although these models are effectively equivalent to the Starobinsky theory for large values of the fields, they are not equivalent for small values.  In particular, one expects large differences in the small-field regime. Therefore, one may correctly identify the range of the validity of the theory by determining its cut-off scale, if  it is considered as an effective field theory. We have discussed  the cut-off  by  looking in the scalar kinetic term, which is similar to kinetic term of an one-dimensional $\sigma$-model.  We have found that,  although the cut-off of the Starobinsky theory is the Planck scale,  for a polynomial function $f(\phi)=\phi^n/M_{\rm p}^{n-2}$ in the general universal attractor model, the cut-off is lower than the inflationary scale for $n<7/3$ (this case  includes also Higgs inflation for $n=2$). However,  the case $n=1$ is particular 
and we have discussed it in more details. In particular, beyond being in agreement with the data, it is valid up to Planckian scales.

  \section*{Acknowledgments}
 We thank J.R. Espinosa for useful discussions.
  A.R. is supported by the Swiss National
Science Foundation (SNSF), project `The non-Gaussian universe" (project number: 200021140236). 
The  research of A.K. was implemented under the ``Aristeia" Action of the 
``Operational Programme Education and Lifelong Learning''
and is co-funded by the European 
Social Fund (ESF) and National Resources.  It is partially
supported by European Union's Seventh Framework Programme (FP7/2007-2013) under REA
grant agreement n. 329083.


\end{document}